# Effect of X-Ray Irradiation on Threshold Voltage of AlGaN/GaN HEMTs with pGaN and MIS Gate

Yongle Qi[1], Suzhen Wu[*]

*Abstract*—Characteristic electrical curves of GaN HEMT devices from Infineon and Transphorm are compared at different X-ray radiation dose. It is shown that the device with pGaN gate is more robust having a stable threshold voltage ($V_{th}$). The $V_{th}$ of device with MIS gate shifts towards negative direction firstly and shifts to forward direction then. A qualitative analysis is performed in the paper. Such dynamic phenomenon is caused by releasing and trapping effects of radiation induced charges both in the dielectric layer and the interface of the device. It is summarized that pGaN gate based GaN HEMT is a promising solution for further space use of electric source.

*Index Terms*—Gallium Nitride, High Electron Mobility Transistor, X-ray Radiation, Threshold Voltage

## I. INTRODUCTION

IN space use, the satellites and their components are required to survive in a harsh radiation environment. In order to be useful in orbit, materials and devices must be able to tolerate damage caused by all kinds of cosmic radiation [1]. The radiation from solar flares is mostly x-rays, gamma-rays, protons and electrons [2]. A fundamental understanding of the reliability and failure mechanisms in these devices is critical to further technology development and commercialization.

In the last few years, top groups from all over the world have extensively studied the effects of $^{60}$Co γ-ray irradiation on conventional depletion-mode (D-mode) AlGaN/GaN HEMTs (Schottky gate of contact of metal and semiconductor)[3]. Significant degradation of AlGaN/GaN HEMTs was observed only after a γ-ray ($^{60}$Co) dose of many tens or even hundreds of Mrad(Si)[4]. Devices show a negative shift in threshold voltage, which is dominated by an increase in trap density. Other experiments [5], [6] and [7] with similar results suggest that damage due to particle irradiation is of much more concern in GaN-based devices, which are more sensitive to displacement damage than ionization effects.

However, there has been little research on the radiation reliability of device with enhancement-mode (E-mode) device with pGaN gate. In the present paper, we compare the x-ray radiation hardness of device with pGaN gate and MIS structure. The mechanism of threshold voltage shift is mainly discussed.

## II. EXPERIMENTS

To gain more insight into the matter of radiation tolerance in GaN-based HEMTs, a radiation-damage study was performed on GaN based HEMT devices. The first device is IGO60R070D1 from Infineon, identified as device type "A". There is a pGaN layer on the gate to realize E-Mode. The second device is TPH3205 from Transphorm, Inc, which is identified as device type "B". It is a D-Mode device of MISHEMT from Transform separated from a cascade structure device. No special structure to improve the vulnerability to the radiation environment was employed for either type of device.

Fig.1 (a) is cross section image of the device A. The pGaN thickness is around 240 nm. There is a pGaN ridge on each side of the drain to suppress the current collapse [8]. The inset is the magnified image of the red frame. There are pGaN ridges on both sides of the drain electrode to suppress current collapse. Fig.1 (b) shows the cross-section structure of device B. The dielectric insulator thickness is around tens nanometers.

X-ray irradiation was performed at a dosage rate of 100rad/s on the device surface as normal. The accumulated dose varies from 100K to 200K to reveal the influence on electrical parameters of the compared devices. It is noted that the I-V curve is measured instantly using Agilent1050B after irradiation.

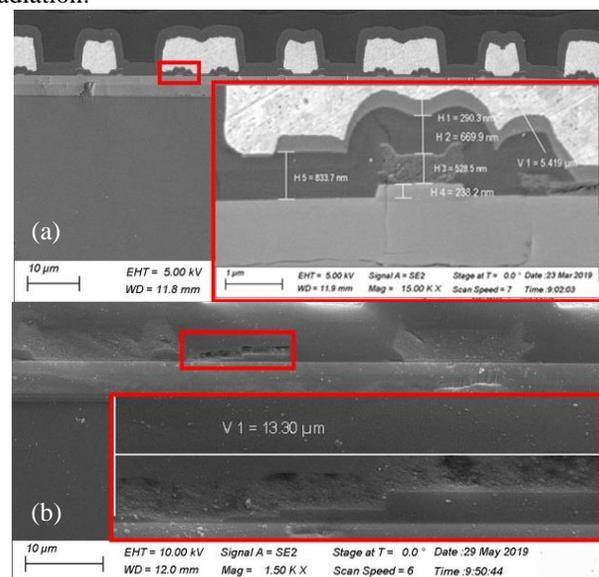

Figure. 1. Cross Section Images of the Devices

Yongle Qi is with Key Laboratory of Microwave and Millimeter Wave Monolithic Integrated and Modular Circuits, Nanjing 210001, China. (e-mail: 875949987@qq.com).

Suzhen Wu is with China electronics technology group corporation No.58 research institute, Wuxi 204035, China. (e-mail: nlgwsz@163.com).

## III. RESULTS AND DISCUSSION

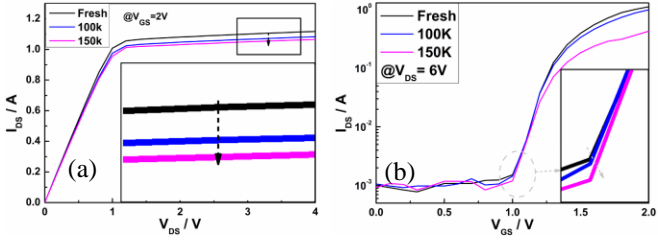

Figure. 2. (a) Output curve and (b) transfer curve of the device with pGaN gate

Fig. 2. shows the output and transfer curves of the device with pGaN gate. The drain current decay and the threshold voltage shift are not obvious. According to the inset of Figure 2(b), the $V_{th}$ shift towards the positive direction slightly which is correspond with the former work. It is reported that the acceptor like traps cause the $V_{th}$ shift [9]. At the same time, during the epitaxy growth process, high density of $H^+$ contamination exits in the epitaxy layer that lowers the activation rate of the acceptor impurities [10]. The inactivated acceptor impurities (Mg ion) might be activated by the ray energy and the $V_{th}$ shifts then.

One proposed explanation for the radiation tolerance of GaN-based HEMTs is that the GaN material itself is so intrinsically defective that creating more defects by irradiation makes little difference [11]. Another proposed explanation is that the threshold energy ($E_d$) for atomic displacement in GaN is higher than in other III-V materials such as GaAs, and that proportionately fewer atoms are displaced [12].

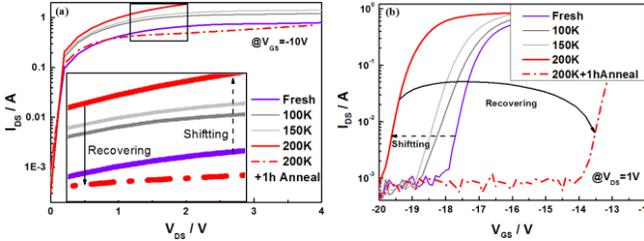

Figure. 3. (a) Output curve and (b) transfer curve of the device with MIS gate

Figure 3 gives the output and transfer curves of the device with MIS gate. It is shown that the drain current increases with the increasing incident dose from 100K to 200K. However, it degrades to a quite low level after 1 hour according to inset of Fig. 3a. Fig. 3b is the transfer curve. Apparently, the $V_{th}$ shift dramatically towards the negative direction with the increasing incident dose at first and then shift positively to -14.5V after 1 hour.

With respect of the MIS HEMT, $V_{th}$ can be expressed as [13]:

$$V_{th} = 2\varphi_F + \varphi_B + \varphi_{ms} - \frac{Q_{ot}+Q_{it}}{C_{ox}} - \frac{Q_m}{C_{ox}} \quad (1)$$

In which, $\varphi_F$ is the potential difference of the middle of the band and the Fermi level, $\varphi_{ms}$ is the contact potential difference of the gate metal and the semiconductor, $\varphi_B$ is the potential difference of the space charge in semiconductor and the insulator, $Q_m$, $Q_{ot}$ and $Q_{it}$ are mobile charges in the insulator, fixed charges and the interfacial charges respectively. $C_{ox}$ is the insulator capacitance which is expressed as:

$$C_{ox} = \frac{\varepsilon_{ox}\varepsilon_o}{d_{ox}} \quad (2)$$

$\varepsilon_{ox}$ and $\varepsilon_o$ are respectively relative dielectric constant and permittivity of free space, $d_{ox}$ is thickness of the insulator. The capacitance is assumed as a constant in the following analysis.

During the X-ray incidence process, energy is transferred to an electron in the valence band by the incoming particle, raising it to the conduction band and creating a corresponding hole in the valence band that causes the production of electron-hole pairs (ionization) [14]. The density of the electron-hole pairs is influenced by the sample quality and the doping level. Fig.4 illustrates the shifting of the irradiation induced excited carriers at electrical stress. A constant negative voltage is applied on the gate that accelerates the excited electrons transferring to the potential well during the irradiation process. Meanwhile, parts of the excited holes are trapped by the deep level in the insulator that forms the positive fixed space charges. It is concluded that it is the irradiation induced positive space charges that make the $V_{th}$ shift negatively. It is a dynamic process influenced by both irradiation and the applied electrical stress on gate. It is the $Q_{ot}$ in the oxide that decreases the $V_{th}$ referred by equation (1).

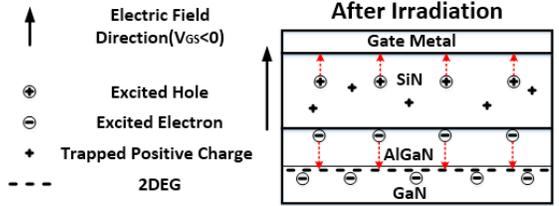

Figure. 4. The irradiation Induced Electron and Hole Pairs Shift Driven by the Applied Gate Voltage

However, after 1 hour recovering (Anneal in air) procedure, $V_{th}$ shifts positively according to Fig. 3(b). GaN HEMTs fabricated under Ga-rich N vacancies were suggested to be responsible for N vacancies and divacancies can be generated during the irradiation. At the operating bias condition, these acceptor-like traps were negatively charged, leading to the positive shift in $V_{th}$. In general, the radiation damage based on positive charge to AlGaN/GaN HEMTs results in mobility degradation and an increase in the threshold voltage, both of which lead to reductions in peak transconductance and drain current [15]. It is noted that the recombination process always happens after experiments of total dose [16]. Totally speaking, it is the irradiation induced negative charges ($Q_{it}$) in the interface that shift the $V_{th}$ positively which correspond with (1).

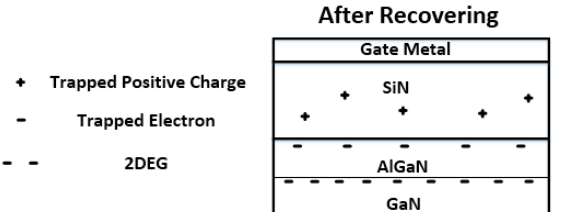

Figure. 5. The Trapped Electrons Reduce the Density of the 2DEG

## IV. CONCLUSION

pGaN gate based GaN HEMT device is more robust to radiation for its high intrinsic defect densities and high threshold energy of GaN material. Regarding MIS HEMT, although the mechanism is still not clear, there are two facts that dominate the threshold voltage shift respectively with time-varying. The negative $V_{th}$ shift might be caused by intrinsic free ions and radiation induced electron-hole pairs in the dielectric. The positive $V_{th}$ shift and current degradation can be modeled by negative trapped charge near 2DEG. The compared results suggest that the GaN HEMT based on pGaN gate is a promising Rad Hard solution for power conversion use in space.